# Dust aerosols above the south polar cap of Mars as seen by OMEGA

Mathieu Vincendon[1], Y. Langevin[1], F. Poulet[1], J.-P. Bibring[1], B. Gondet[1], D. Jouglet[1], OMEGA Team

[1]Institut d'Astrophysique Spatiale, CNRS/Université Paris Sud, Bâtiment 121, 91405 Orsay, France (mathieu.vincendon@ias.u-psud.fr)

**Abstract:** The time evolution of atmospheric dust at high southern latitudes on Mars has been determined using observations of the south seasonal cap acquired in the near infrared (1–2.65 µm) by OMEGA/Mars Express in 2005. Observations at different solar zenith angles and one EPF sequence demonstrate that the reflectance in the 2.64 µm saturated absorption band of the surface $CO_2$ ice is mainly due to the light scattered by aerosols above most places of the seasonal cap. We have mapped the total optical depth of dust aerosols in the near-IR above the south seasonal cap of Mars from mid-spring to early summer with a time resolution ranging from one day to one week and a spatial resolution of a few kilometers. The optical depth above the south perennial cap is determined on a longer time range covering southern spring and summer. A constant set of optical properties of dust aerosols is consistent with OMEGA observations during the analyzed period. Strong variations of the optical depth are observed over small horizontal and temporal scales, corresponding in part to moving dust clouds. The late summer peak in dust opacity observed by Opportunity in 2005 propagated to the south pole contrarily to that observed in mid spring. This may be linked to evidence for dust scavenging by water ice-rich clouds circulating at high southern latitudes at this season.

## 1. Introduction

The optical depth of dust aerosols is a critical parameter for global climate models (Forget et al., 1999). The behavior of dust aerosols on Mars is complex, depending on the location, the season, and the year. Simulations have been recently carried through so as to better understand the martian dust cycle (Newman et al., 2002a, 2002b; Kahre et al., 2006), notably in the south polar regions of Mars (Toigo et al., 2002).

Recent missions have improved our comprehension of the temporal and spatial variations of the optical depth of dust aerosols. The two Mars Exploration Rovers (MERs) have performed direct measurements of the optical depth in the visible via sunlight extinction measurements on a daily time scale at each landing site (Lemmon et al., 2004;

Lemmon et al., 2006). Clancy et al. (2003) and Wolff and Clancy (2003) have analyzed Emission Phase Function (EPF) sequences acquired by the Thermal Emission Spectrometer (TES). They obtained the seasonal and geographical distribution of the optical depth in the visible, the particle size and the type (dust or water ice) of aerosols for latitudes ranging from 45° S to 45° N. The Mars Orbiter Laser Altimeter (MOLA) has mapped the densities of absorptive and reflective clouds during two Mars years (Neumann et al., 2003) and the Mars Orbiter Camera (MOC) has obtained the distribution of dust storm during southern winter and spring of 1999 (Cantor et al., 2001). The mapping of the optical depth of dust aerosols over most of the planet has been performed by TES during three years at 9 µm. The major global seasonal patterns have been determined (Smith, 2004, 2006). The dust opacity is not indicated in (Smith, 2004) and (Smith, 2006) for latitudes and times corresponding to ice-covered regions. However, a few maps of the optical depth of dust aerosols in mid-spring are provided in Kieffer et al. (2000) on the basis of TES observations. The optical depth in the UV has been derived from the SPICAM nadir dataset as a secondary product of the ozone analysis (Perrier et al., 2006).

The atmosphere of Mars interacts in a complex manner with the polar caps. Deposition and removal of dust on ice-covered regions is observed (Kieffer, 1990; Langevin et al., 2005). Such processes could be linked with frost sublimation and deposition via nucleation processes (Gooding, 1986). The development of the cryptic region of the south seasonal cap (Kieffer et al., 2000) is now clearly linked to surface dust contamination processes (Langevin et al., 2006; Kieffer et al., 2006). The role of atmospheric dust deposition with respect to sub-surface venting remains to be clarified.

The visible and near-IR imaging spectrometer OMEGA (Observatoire pour la Minéralogie, l'Eau, les Glaces, et l'Activité) has achieved an extensive coverage of south polar regions of Mars since early 2004, providing significant results in the analysis of the seasonal evolution of surface properties such as ice composition, dust contamination of the ice, and the mean distance between scattering interfaces which is linked to grain size (Bibring et al., 2004; Douté et al., 2007; Langevin et al., 2007). Aerosols significantly modify reflectance spectra of Mars obtained in the near-IR (Drossart et al., 1991; Erard et al., 1994). Above the south polar cap of Mars, suspended dust increases the reflectance in saturated bands of surface $CO_2$ ice (Langevin et al., 2006).

We have retrieved the contribution of aerosols in the OMEGA dataset acquired over the south seasonal cap from mid-southern winter to late-southern summer of martian year 27 (January to December 2005). The spatial resolution ranges from 1 to 10 km and the timescale between two observations that partially overlap is frequently between 0.5° and 1° of $L_s$. We use the Monte-Carlo based model of radiative transfer and the properties of airborne particles described by Vincendon et al. (2007a). The method used in this study relies on the assumption that aerosols are mainly free of water ice. The minor role of characteristic absorption bands of water ice at 1.5, 2 and 3 µm demonstrates that this is the case for most OMEGA observations of the southern seasonal cap except close to the southern spring equinox (Langevin et al., 2007). This conclusion is supported by observations in the thermal infrared (Smith, 2004, 2006), and model predictions (Forget et al., 1999; Montmessin et al., 2004). This study will therefore focus on the spatial and temporal variations of dust aerosols.

In Section 2, we present the dataset and the method used to model the contribution of dust aerosols. In Section 3 we perform the mapping of the optical depth of aerosols at 2.64 μm over most places of the seasonal cap from $L_s$=220° to $L_s$=290° and we determine the optical depth above the perennial cap on a longer time scale ranging from $L_s$=185° to $L_s$=340°. Section 4 presents a discussion of the results and their implications.

## 2. Aerosols modeling

### 2.1. Radiative transfer model and scattering properties of aerosols

The modeling of the contribution of atmospheric dust between 1 and 2.65 μm is performed using the Monte-Carlo based model of radiative transfer and the properties of airborne particles presented in Vincendon et al. (2007a). The model simulates the path of photons in a layer of identical particles and the interaction with the surface which is supposed to be Lambertian. A constant Henyey–Greenstein (H–G) phase function with an asymmetry parameter of 0.63 has been selected for aerosols between 1 and 2.6 μm according to Ockert-Bell et al. (1997). A single scattering albedo which range from 0.971 to 0.976 has been inferred from OMEGA observations of optically thick aerosols (a constant value of 0.974 is used for simplicity in this study). With these assumptions, the reflectance factor (RF, defined by $I/F/\cos(i)$) depends only on the photometric angles and on two wavelength dependent parameters: the optical depth of aerosols $\tau(\lambda)$ and the Lambert albedo of the surface $A_L(\lambda)$. Vincendon et al. (2007a) have demonstrated that weak variations of the phase function with wavelength can be compensated by weak variations of the optical depth for the moderate phase angle variations considered here. The influence of the preceding assumptions on the retrieved values of the optical depth will be discussed in Section 3.3.

Following the approach of Vincendon et al. (2007a), we have populated a look-up table of modeled reflectance factors as a function of the photometric angles (incidence, emergence, and azimuth), the Lambert albedo of the surface and the optical depth of aerosols. This table is used in the next sections to model and/or remove the contribution of aerosols in observations. This model has been previously used only for observations acquired in the nadir pointing configuration. Several OMEGA observations at high southern latitudes are obtained with emergence angles significantly different from 0. Moreover, several spot pointing observations (i.e. Emission Phase Function, EPF) have been acquired by OMEGA. We have validated our Monte-Carlo based model for off-nadir configurations by comparing its results with the algorithm of Forget et al. (2008) which provides results similar to that of the SHDOM code (Evans, 1998) for low optical depths (Fig. 1).

A detailed study of the contribution of water ice aerosols to the OMEGA dataset acquired above the south polar cap is provided by Langevin et al. (2007). This study is based on the water ice absorption bands at 1.5, 2 and 3 μm. Strong water ice clouds signatures are detected in the outer regions of the cap during early spring ($L_s$~190°) and predicted by GCM simulations at that time (Montmessin et al., 2004). From mid-spring to mid summer most OMEGA observations are nearly free of water ice either as aerosols or on the surface. The weak water ice signatures observed above the perennial cap from $L_s$=300° are attributed to the surface. We have limited our study of dust aerosols to observations for which the contribution of water ice aerosols can be neglected. This is the case for the major part of the OMEGA dataset obtained in 2005 above the southern seasonal cap.

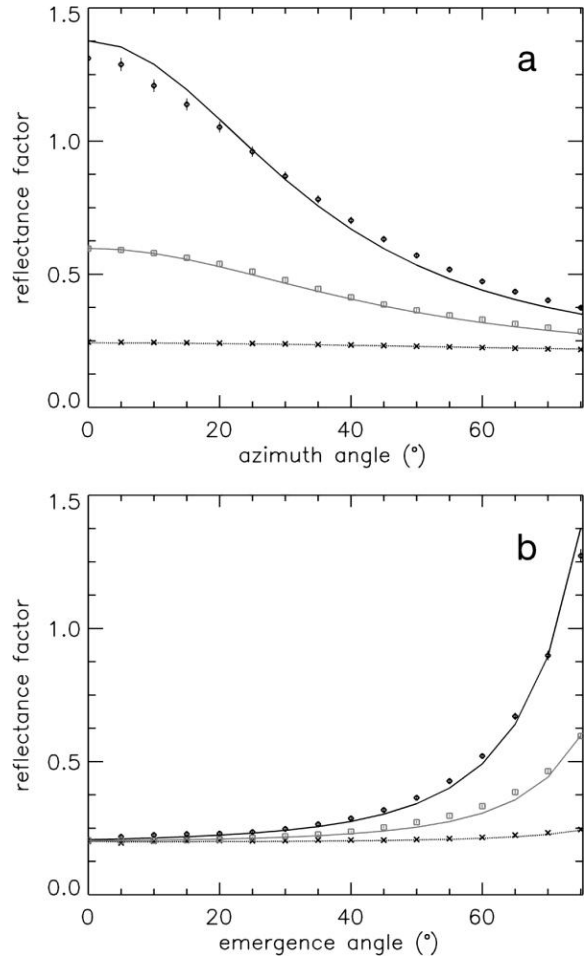

**Figure 1.** *Comparison between the algorithm of Forget et al. (2008) and our Monte-Carlo model for an optical depth of 0.16, a surface albedo of 0.2, a single scattering albedo of 0.974 and a Henyey–Greenstein phase function with an asymmetry parameter of 0.63. Variations of the reflectance factor (I/F/cos(i)) with the azimuth angle (a) and the emergence angle (b) are shown for 3 values of the solar incidence angle i (i=72°: black solid line for the model of Forget et al. and diamonds for the Monte-Carlo model; i=66°: resp. gray solid line and squares; i=25°: resp. dashed line and crosses). The error bars for the Monte-Carlo model (±2% for $3\times10^7$ simulated photons) are indicated. The two models give similar results.*

## 2.2. Spectral characteristics of dust aerosols at high southern latitudes

The variation of the optical depth with the wavelength has been previously studied with our method in the northern hemisphere during summer (Vincendon et al., 2007a), providing a factor of 2.6 between the optical depths at 1 and 2.5 μm and in the Meridiani region (Vincendon et al., 2007b; factor of 1.9 between 1 and 2.65 μm). In both cases the optical depth monotonically decreases between 1 and 2.5 μm. We discuss here the spectral-variations of the aerosol optical depth at high southern latitude regions during mid winter and in early summer.

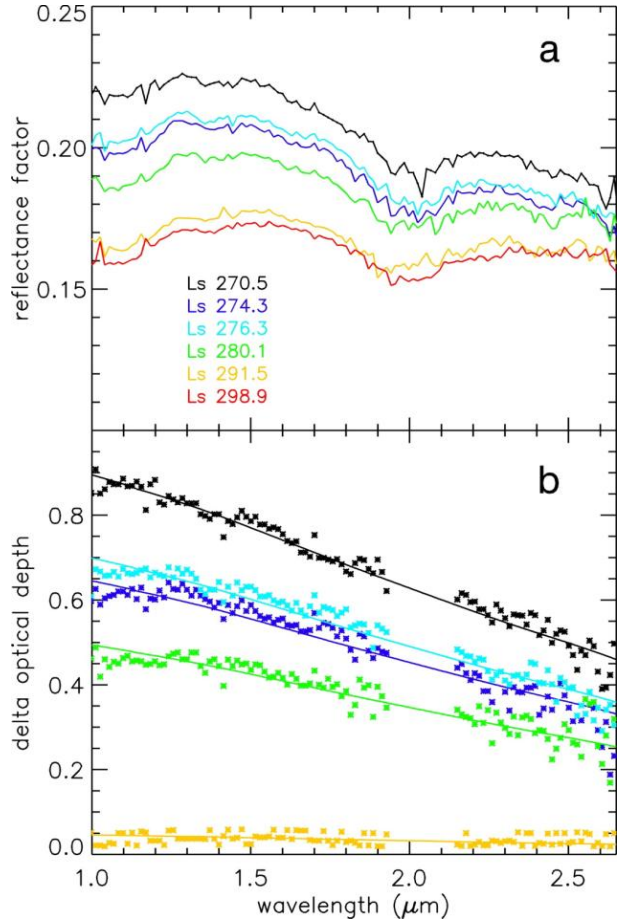

**Figure 2.** *(a) Observation of a dark region (75.4° E, 77.2° S) during a period of decrease of the optical depth of aerosols ($L_s$=270°–300°) with similar photometric angles (nadir, i~54°). (b) Retrieval of the wavelength dependence of the difference of optical depth between the red spectrum of figure (a) and the others spectra at higher optical depth. The same dependence of the optical depth with wavelength applied to different optical depths at 1 μm (thick solid lines) provides satisfactory fits of the 5 observations between $L_s$=270° and $L_s$=292° presented in (a) (crosses, with the same color code). The region near 2 μm which corresponds to the major atmospheric $CO_2$ absorptions is not indicated.*

After the southern summer equinox, TES observes decreasing dust opacities at high southern latitudes (Smith, 2004). A sequence of six observations of a dark region at 75.4° E, 77.2° S has been obtained between $L_s$=270.5° and $L_s$=298.9° with similar photometric angles (nadir pointing, solar incidence angle of 54°), which minimizes the possible impact of surface photometric properties. These observations are presented in Fig. 2a. They show that the reflectance factor strongly decreases above dark surfaces during this period, which is consistent with a decrease in the dust optical depth. We consider as a reference the observed spectrum at $L_s$=298.9° (red spectrum in Fig. 2a), as it corresponds to the minimum dust opacity observed by TES. From these observations, it is possible to retrieve the delta optical depth at each wavelength for the higher aerosol loadings using the approach presented in Vincendon et al. (2007a). The retrieved optical depth increments for the five observations with higher dust loadings are then fitted by a linear combination of optical depth functions for different particle sizes computed with Mie theory (Clancy et al., 2003, Fig. 13). A value of 1.9±0.2 for the ratio $\tau(1\ \mu m)/\tau(2.65\ \mu m)$ is consistent with the whole set of observations (Fig. 2b).

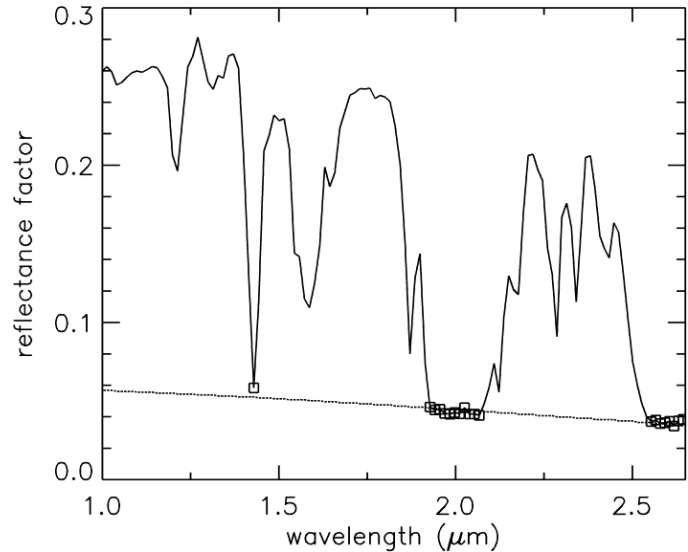

**Figure 3.** *Reflectance spectrum of translucent slab CO2 ice (mean spectrum of a region at 60° S and 0° E between Ls=136° and Ls=147°). The solar incidence is not, vert, similar80°. The emergence is not, vert, similar0° (nadir). The dashed line corresponds to the contribution of aerosols to the reflectance factor for a perfectly absorbing surface assuming the same dependence on wavelength of the optical depth as that presented in Fig. 2.*

Constraints on the spectral properties of aerosols at southern latitudes in mid winter ($L_s$=130°–150°) can be obtained from observations of a translucent slab of $CO_2$ ice with saturated absorption features at 1.43, 2 and 2.6 μm observed in the seasonal cap with OMEGA at 0° E, 60° S (Langevin et al., 2006). A typical example of a translucent slab ice spectrum observed by OMEGA is shown in Fig. 3. The signal in the three saturated bands is due to scattering events before absorption by the surface ice (Douté et al., 2007). Scattering events can occur in the aerosol layer or at the surface of the ice if it is rough or dust contaminated. The solar incidence angle is >80°, hence the contribution of scattering by aerosols is large. An upper boundary of the optical depth in saturated bands at 1.43, 2 and 2.6 μm can be obtained by assuming that this contribution dominates. The reflectance factor in saturated bands can be adequately modeled (dashed line in Fig. 3) by an aerosol contribution with a wavelength dependence as determined at $L_s$~270°–300° from Fig. 2 and a low optical depth of 0.1 at 2.6 µm. This result supports the assumption that the wavelength dependence of the optical depth presented in Fig. 2 is appropriate for aerosols above the south polar cap from mid-winter to mid-summer as a first approximation.

The interpretation of the observed decrease of the backscattered light by a factor of 1.9 between 1 and 2.65 µm is not straightforward. On the basis of Mie theory, this spectral dependence would require a mean particle size smaller than the value of ∼1.6 µm obtained by different authors (see, e.g., Wolff et al., 2006). This is consistent with the smaller mean size (∼1.2 µm) derived by Drossart et al. (1991) from IRTM near-IR observations. Other parameters may also play a role, such as the shape of the particles (Mishchenko et al., 1996), a size distribution depending on altitude (Korablev et al., 1993; Montmessin et al., 2006) or a bimodal size distribution (Chassefiere et al., 1995; Montmessin et al., 2002). A review on that subject has been provided by Korablev et al. (2005). Vincendon et al. (2007a) have derived a significantly steeper decreasing slope of backscattered light at high northern latitudes during summer using a similar set of observations as in Fig. 2. This is consistent with observations by Clancy et al. (2003) of a smaller mean particle size of aerosols (1 µm) at that time in the northern hemisphere.

### 3. Mapping of the optical depth of aerosols

#### 3.1. Contamination of surface ice with dust

Most $CO_2$ spectra observed above the south polar cap by OMEGA have a saturated band at 2.6–2.65 µm (Langevin et al., 2007). The signal in this band never reaches zero, mainly due to scattering by dust particles before absorption by $CO_2$ ice. This dust can be either above the surface (aerosols) or at the surface. We can discriminate between surface dust and atmospheric dust using observations with different geometries acquired during a period of stable atmospheric dust opacity and constant surface properties (Vincendon et al., 2007a). Series of observations with significant variations of the solar incidence angle over short time periods have been obtained by OMEGA around the summer solstice and for latitudes between 70° S and 80° S. In Fig. 9 of Langevin et al. (2007), the region at 316° E, 78° S ($L_s$∼263°) is observed with two incidences (56° and 66°). The spectrum at 66° is the average of 2 spectra acquired respectively less than 1° of $L_s$ before and after that at 56° to minimize effects of possible variations of the optical depth during the period. The reflectance of the reconstructed surface spectrum is ∼0 in the saturated $CO_2$ ice band at 2.64 µm. We also observe that the surface reflectance at 2.64 µm is ∼0 at three other places and times where two consecutive observations ($L_s$ gap of 0.18° to 0.35°) with different photometric angles have been obtained by OMEGA: at 0° E, 80° S, $L_s$=263° (Fig. 4a), at 280° E, 80.4° S, $L_s$=250° (Fig. 4b) and at 284° E, 73.2° S, $L_s$=241.7° (Fig. 4c). One Emission Phase Function sequence (EPF, during which the same surface element is observed for most of the visibility window on a given orbit) has been obtained with OMEGA above surface $CO_2$ ice. The observation takes place above the permanent cap at $L_s$=288°. The observed region slightly varies during the sequence (from 85.7° S to 86.4° S and from 326° E to 4° E). The evolution of the photometric angles during the sequence is indicated in Fig. 5a. The reflectance factor strongly increases with emergence and phase angle (from 4 to 40%, see Fig. 5) whereas it would have been constant for a Lambert surface without aerosols. A very good fit of the variation of the reflectance factor is obtained for a Lambert surface with an albedo of ~0.7% and with an optical depth of 0.17 (Fig. 5b). These observations demonstrate that in most cases the contribution of surface dust contamination is very low compared to that of aerosols, which makes it possible to retrieve the optical depth directly from the reflectance factor in saturated $CO_2$ ice bands.

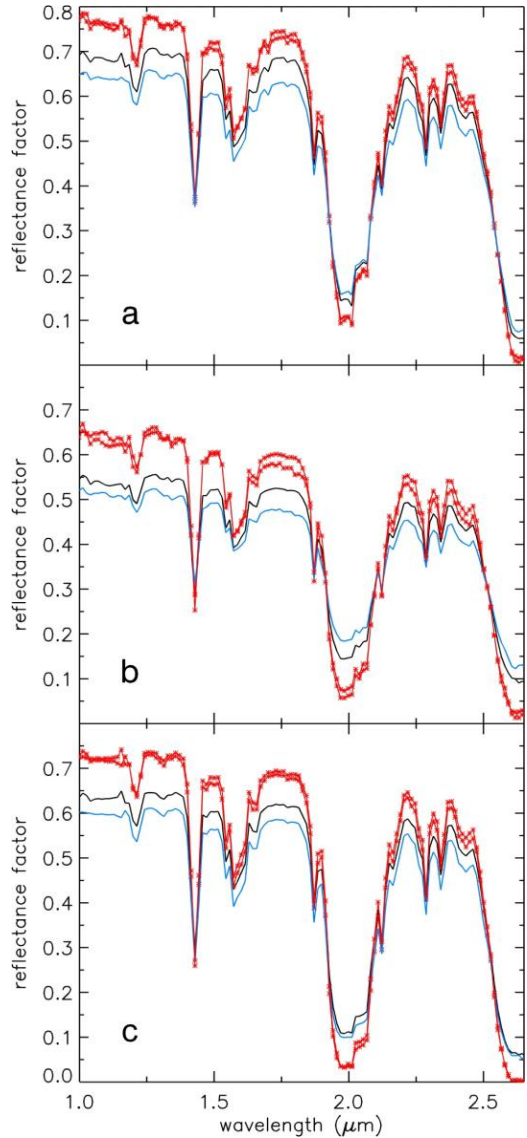

**Figure 4.** *Observed variations of $CO_2$ ice reflectance spectra with photometric angles (black and blue solid lines) and reconstructed surface spectra (red stars) for different regions (a, 0° E, 80° S; b, 238° E, 80° S; c, 284° E, 73.2° S), solar longitudes (a, 263°; b, 250°; c, 241.7°) and photometric angle variations (a, black: i=59°, e=0° and blue: i=70°, e=0°; b, black: i=68°, e=12°, ϕ=66° and blue: i=75°, e=35°, ϕ=68°; c, black: i=68°, e=8°, ϕ=61° and blue: i=65°, e=26°, ϕ=65°). The time intervals between observations are small (a and c, 0.6 days; b, 0.3 days). The reflectance factor of the surface is <1% at 2.64 µm for the three observations.*

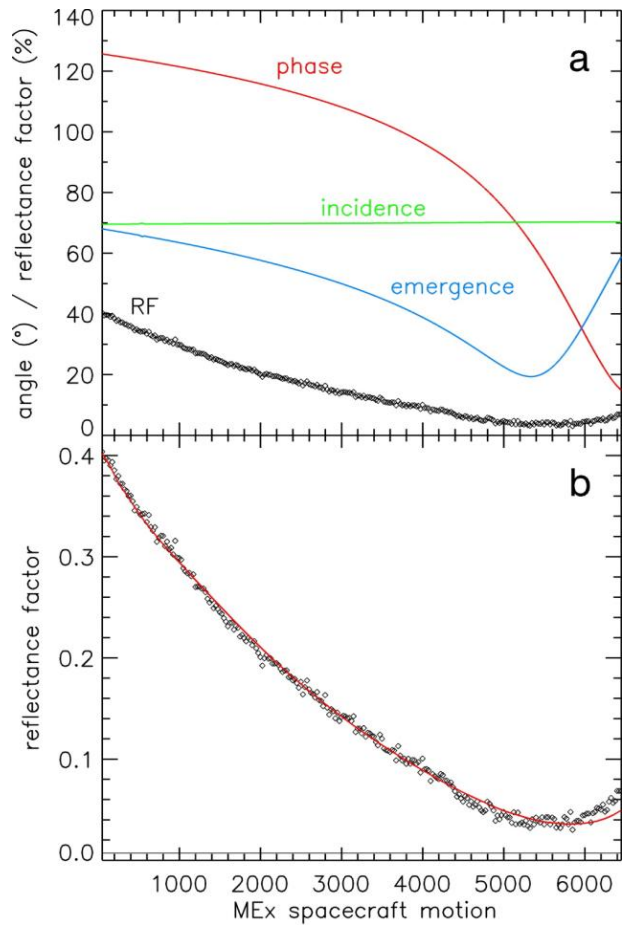

**Figure 5.** *EPF sequence above the south permanent cap (~86° S, ~345° E) at $L_s$=288°. (a) Variation of the photometric angles during the sequence as a function of time. The reflectance factor (diamonds) at 2.64 μm (saturated $CO_2$ band) undergoes strong variations from 40 to ~4%. (b) The Monte-Carlo model provides a satisfactory fit (thick red solid line) of the Reflectance Factor (diamonds, same as (a)) for a Lambert surface of 0.7% (thin solid line) and an optical depth of 0.17. The mismatch for phase angles smaller than 20° could be due to an increase of the contribution of the surface in the anti-solar direction.*

This assumption is clearly not valid in specific places of the seasonal cap and during specific periods. The cryptic region discovered by Kieffer et al. (2000) and widely studied in the near-IR by Langevin et al. (2006) extends from 60° E to 210° E at latitudes around 80° S between $L_s$=200° to $L_s$=230°. High levels of surface dust contamination of the surface ice have been inferred in this region from spectral modeling (Langevin et al., 2006). A representative spectrum of this region is shown in Fig. 6 (green spectrum). Sub-pixel spatial mixing of ice-covered and ice-free surfaces is observed close to the sublimation front (Langevin et al., 2007) and for some patches inside homogeneous regions of clean $CO_2$ ice (Fig. 6, black spectrum). These two situations result in a similar spectral signature in the wavelength range of interest (low reflectances in the continuum and high reflectances in the absorption bands), which is characteristic of sub-pixel spatial mixing at either a microscopic or macroscopic scale. The reflectance at 5 μm gives an indication on the mean temperature inside the pixel and makes it possible to discriminate between dust at the ice temperature and warmer dust (see Langevin et al., 2007).

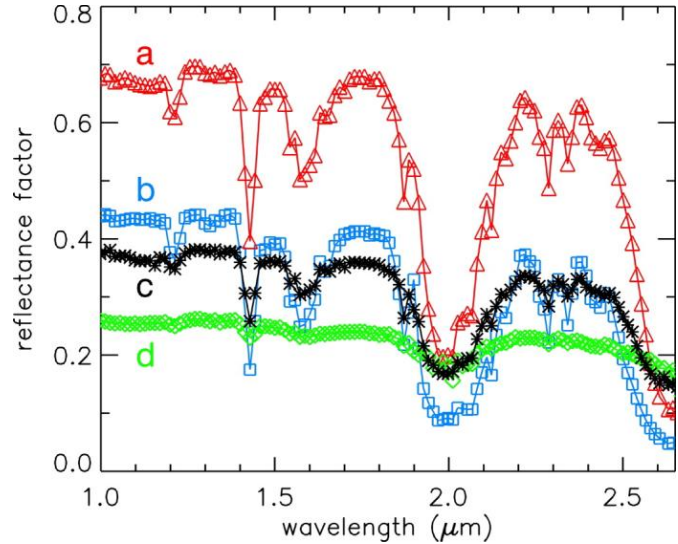

**Figure 6.** *Typical reflectance spectra of $CO_2$ ice with different types of dust contributions. (a) Red triangles: bright region with a major contribution of atmospheric dust (region analyzed in Fig. 16). (b) Blue squares: intra-mixture of a few 10 ppm of dust grains within ice observed through a clean atmosphere (268.5° E, 73.4° S and $L_s$=222.2°). (c) Black stars: spatial mixing of ice-covered and ice-free regions (300.6° E, 79.4° S and $L_s$=256.5°). (d) Green diamonds: $CO_2$ ice contaminated by surface dust in the cryptic region (Langevin et al., 2006, Fig. 3).*

### 3.2. Methodology for mapping the aerosol optical depth over the ice cap

Few series of closely spaced OMEGA observations with different geometry have been obtained. They demonstrated that when major $CO_2$ ice bands are strong the contribution of surface dust to the reflectance factor is very small (<1%). The atmospheric $CO_2$ gas absorption at 2.7 μm has no impact below 2.65 μm and the absorption band of $H_2O$ gas at 2.6 μm (Encrenaz et al., 2005) has negligible impact for our method (see Section 3.3). The observed flux at 2.6–2.65 μm can therefore be used to map the total optical depth of aerosols for each observation made above regions with strong $CO_2$ ice absorptions. It can be determined as an average between the reflectance factors measured by OMEGA at 2.631 and 2.644 μm. The reflectance factor varies monotonically as a function of the optical depth for a given set of photometric angles. Therefore, the optical depth can be unambiguous determined by comparing the observed reflectance factor at 2.6–2.65 μm with the look-up table resulting from the Monte-Carlo model for the appropriate set of photometric angles.

We have performed this procedure for OMEGA observations acquired during southern spring and summer of 2005. Before $L_s$=220°, the contamination of the ice spectra by surface and atmospheric water ice is not negligible except above the perennial cap (Langevin et al., 2007). After $L_s \sim 290°$, the fraction of the surface covered by $CO_2$ ice is restrained to the small perennial cap and OMEGA has performed narrow tracks with high spatial resolution due to the decrease of the altitude of Mars Express. Maps have therefore been obtained between $L_s=220°$ and $L_s=290°$, and the optical depth above the perennial cap has been derived from $L_s=185°$ to $L_s=340°$.

The solar incidence angle is lower than 80° and the emergence angle is lower than 40° during this period. This is compatible with the plan parallel hypothesis of our Monte-Carlo model. Most incident photons scattered in the layer of aerosols are distributed over an area <1/2° of latitude for $i=80°$.

We have mentioned in Section 3.1 that significant surface dust contamination of the $CO_2$ ice spectrum is detected in some places and times of the cap. Our method results in an overestimation of the optical depth above these regions of low albedo in the continuum. The reflectance level in the continuum can however not been used alone for the removal of inadequate pixels: during mid-spring, large regions of the cap appears have both a low reflectance at 1 µm, and a low reflectance at 2.6 µm (Fig. 6, blue spectrum). These regions correspond to an intra-mixture of small dust grains within the ice layer: photons at 2.6 µm are absorbed before being scattered by the dust grains (Shkuratov et al., 1999 ; Poulet et al., 2002). In this case the reflectance at 2.6 µm can still be used to determine the optical depth despite the low albedo the continuum. The selection of pixels which can be used to map the optical depth is therefore not straightforward.

A method for selecting pixels free of dust contamination has been derived from the relationship between the observed reflectance factor at 1.08 µm and the optical depth modeled from the reflectance at 2.6 µm. When there is optically thick dust partially covering the surface, its reflectance factor at 1.08 and 2.6 µm are similar, and they combine linearly with the high reflectance factor at 1.08 µm and very low reflectance factor at 2.6 µm of $CO_2$ ice. For aerosols, single scattering events dominate, which favors high emergence angles, away from the nominal nadir looking geometry of OMEGA. Therefore, the relationship between the continuum and the 2.6 µm region is different from that relevant for surface dust contamination as demonstrated in Fig. 7 for four observations. For a wide range of actual surface albedos, the M-C model predicts a weak relationship between the albedo at 1.08 µm and the optical depth of aerosols (black dashed lines in Figs. 7b–7d). An apparent optical depth can also be modeled when dust partially covers the surface, in which case it is overestimated. The dependence of the apparent optical depth on the albedo at 1.08 µm is much stronger when surface dust is present (red lines in Fig. 7). As can be expected, points following the relationship expected for surface dust correspond in all four cases to relatively low albedos and relatively high modeled optical depths. We have selected a cubic polynomial (Fig. 7, black line), which adequately separate points contaminated by surface dust (top left) from points where the optical depth can be modeled with confidence (bottom right) for the full set of observations of the seasonal cap. This selection criterion has been used when preparing the maps of the optical depth over the ice cap which are presented in Section 4.1.

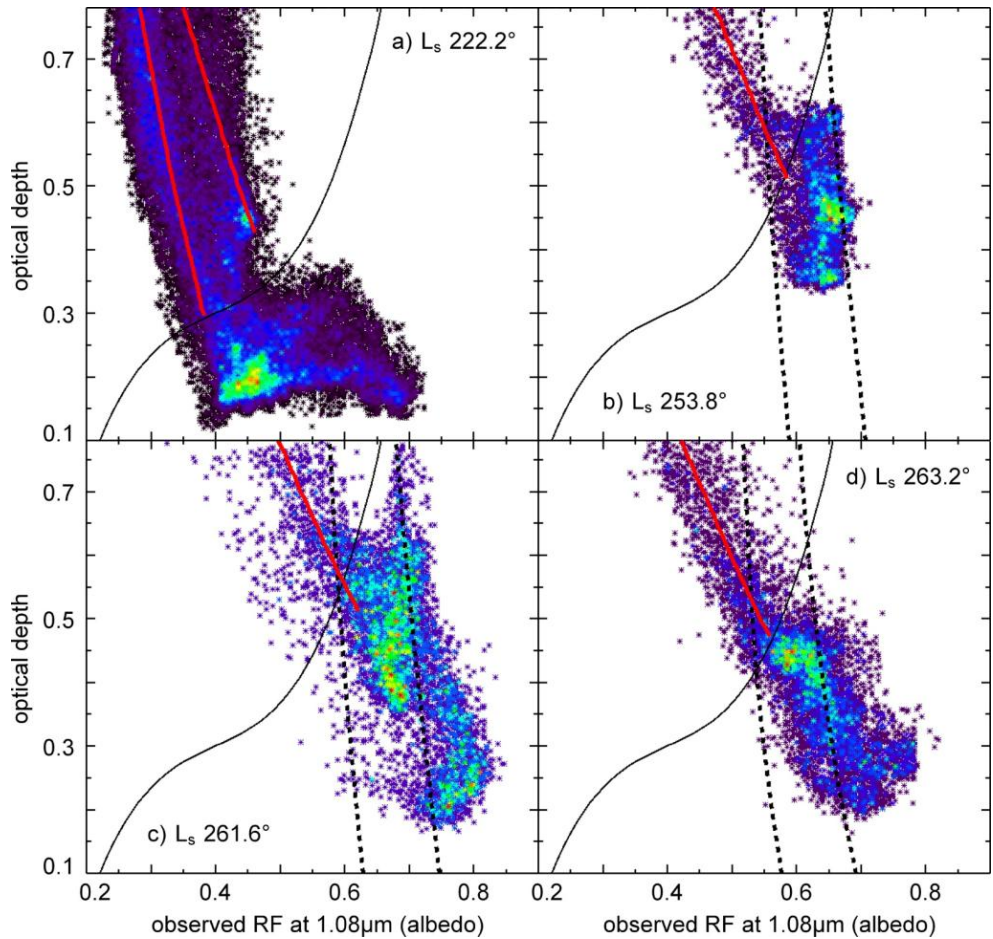

**Figure 7.** *Modeled optical depth as a function of the observed albedo for four observations (a, $L_s$=222.2°; b, $L_s$=253.8°; c, $L_s$=261.6°; d, $L_s$=263°). The rainbow scale indicates the density of points from purple to red. The relationship predicted by the M-C model if there is no contribution of surface dust at 2.6 µm is indicated as a dotted line for (b) i=58°, $A_L$=0.6 and 0.72; (c) i=58°, $A_L$=0.64 and 0.76; (d) i=68°, $A_L$=0.6 and 0.72. At $L_s$=222.2° (a), the observed albedos at 1.08 µm are lower than for other observations even for regions with a very low reflectance at 2.6 µm (hence free of surface dust contamination). The range of observed albedos at 1.08 µm then corresponds to different low levels of dust contamination within the ice (intra-mixture) as discussed in Bibring et al. (2004). The relationship predicted by the M-C model for partial coverage by surface dust within a pixel is indicated as red solid lines. The albedo of dust-covered areas (~0.2) is derived from that of nearby ice-free regions. A reasonable fit for points with high apparent optical depths can be obtained with a single value for the albedo at 1.08 µm of the ice-covered regions for plates (b) and (d) (~0.75) and plate (c) (~0.8). Two mixing lines with low values of ice albedos (0.45 and 0.6) are required to cover the same range in plate (a). The black line corresponds to the polynomial function which has been defined so as to separate regions contaminated by surface dust (top left) from regions where the aerosol optical depth can be reliably derived (bottom right).*

### 3.3. Uncertainties and influence of the assumptions

The noise of the OMEGA instrument is dominated by the read noise (1.85 DN). For most OMEGA observations, the signal-to-noise ($S/N$) ratio is above 100 in the C-channel (Bibring et al., 2004). However we consider here specific observations with a low signal: the surface ice does not reflect light at 2.64 µm and the solar flux is significantly reduced compare to 1 µm. The signal at 2.64 µm strongly varies in our set of observations, depending on the optical depth of aerosols and on the photometric angles. At high altitudes, the drift rate in nadir pointing mode is very slow, and successive scans (2 or 4) are summed so as to avoid oversampling. This increases the $S/N$ by a factor of 1.4 to 2. The $S/N$ is greater than 20 for most observations, reaching 100 for summed observations of regions with high optical depth. It can be lower than 10 for very low optical depth without summation. The relationship between the observed reflectance and the modeled optical depth is nearly linear for small variations of the reflectance (<10%). The noise level therefore results in an uncertainty smaller than ±5% for most observations, with an upper limit of ±10% for some observations at very low optical depths.

A Henyey–Greenstein phase function with an asymmetry parameter of 0.63 has been selected for our wavelength range on the basis of the study of Ockert-Bell et al. (1997). It is useful to assess the impact of a different hypothesis on the shape of the phase function. Therefore, we have performed similar Monte-Carlo simulations with the phase function obtained by Tomasko et al. (1999) at 0.965 µm (Fig. 8, black squares). A single scattering albedo of 0.978 has to be selected with this phase function so as to be consistent with OMEGA observations of dust storms (see Vincendon et al., 2007a). Using the Tomasko et al. phase function results in a scaling of the optical depth values by a factor of 1.22±0.03 for the considered geometries (solar incidence angles between 55° and 75°, viewing geometry close to the nadir pointing). We have performed the same test for two other couples of parameters. Using the phase function of Tomasko et al. and their single scattering albedo (0.937) results in a greater mean multiplying factor (1.36) with a greater dispersion (±0.1) (Fig. 8, green triangles). One must however notice that this lower single scattering albedo leads to an unrealistically low value of the reflectance factor of a dust storm at 0.965 µm (RF=27%). We have also tested the Henyey–Greenstein phase function with a lower single scattering albedo (0.966), which corresponds to the assumption that OMEGA overestimates by 10% the reflectance of dust storms. In this case we obtain a multiplying factor of 1.02±0.005 (Fig. 8, red diamonds).

From this analysis, the uncertainties on the optical properties of aerosols will result in the multiplication of all the optical depth maps by a nearly constant factor ranging from 1 to 1.36, which gives an estimate on the absolute uncertainty of the retrieved optical depth. The observed spatial and temporal variations are very reliable as they depend only weakly on optical parameters over a wide range of hypothesis.

In previous studies of ice-covered regions (Vincendon et al., 2007a), we considered that surface scattering is Lambertian. In this study, the Lambertian hypothesis has no influence on the retrieved optical depth maps as the surface is assumed to be totally absorbing at 2.64 µm. The assumption that the surface reflectance is 0 (when it can reach ∼1% due to surface roughness for clean $CO_2$ ice) results in an overestimation of aerosol optical depth by at most 0.05. The absorption band of $H_2O$ gas at 2.6 µm in OMEGA spectra has been extensively studied (see, e.g., Encrenaz et al., 2005). The amount of water vapor in the atmosphere varies during spring and summer at high southern latitudes, reaching a maximum between $L_s$=270° and $L_s$=330° (Smith, 2004, 2006). The observed band strength varies with a maximum of 10% at the center of the band. The maximum band strength is only 5% at 2.64 μm, which would result in a factor of at most 0.95 between the actual and observed reflectance at this wavelength. Furthermore, only photons scattered by aerosols are observed over ice-covered regions, which reduces the mean column density of water vapor by a factor of 0.6 as the corresponding scale height is half of that of aerosols (11 km in well-mixed conditions). As a consequence the optical depths of aerosols at 2.64 μm are underestimated by at most a few % due to the $H_2O$ gas band centered on 2.6 μm.

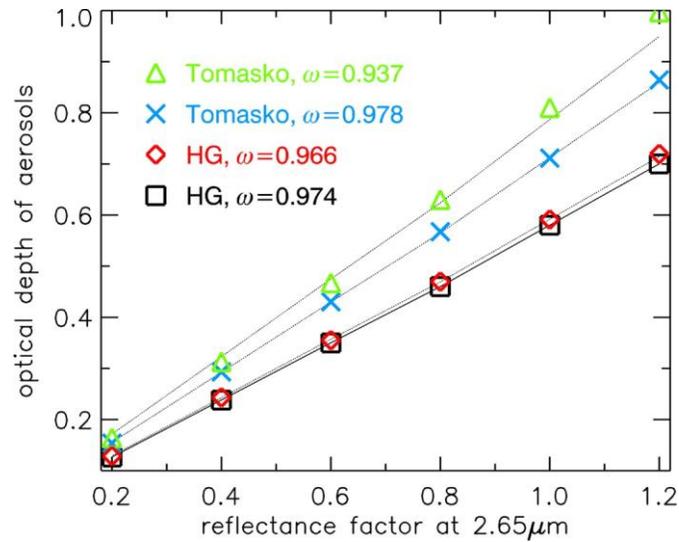

**Figure 8.** *Influence of the selected aerosols parameters on the retrieved optical depths. The black squares correspond to the optical depth as a function of the reflectance factor observed at 2.64 μm for a solar incidence angle of 65°, a nadir viewing geometry and the set of aerosol optical parameters used in our study. Using the phase function of Tomasko et al. (1999) (blue crosses) and/or another value for the single scattering albedo (red diamonds and green triangles) results in the multiplication of the modeled optical depths by a nearly constant factor (dashed lines): 1.02±0.005 for red diamonds, 1.23±0.01 for blue crosses and 1.36±0.05 for green triangles. (For interpretation of the references to color in this figure legend, the reader is referred to the web version of this article.)*

### 3.4. Correlation with photometric angles

Light scattered by aerosols strongly depends on the geometry of observation. In particular, the total column density of grains encountered by photons is proportional to $1/\cos(i)$ (where $i$ is the solar incidence angle). The emergence and incidence angles vary over a wide range in OMEGA observations. We have checked the correlations of observed optical depths on incidence and emergence angles in Fig. 9. As expected, there is no correlation with emergence angle which strengthens the case for our method (Fig. 9a). Low optical depths ($\tau(2.64\ \mu m)=0.2–0.3$) are retrieved for all incidence angles in the range of interest (Fig. 9b). The lack of large optical depths (>0.3) for high incidence angles can be attributed to a trend with latitude and elevation. Very similar diagrams exhibiting no obvious correlation are also obtained if the phase function of (Tomasko et al., 1999) is used. The only case when a possible dependence of the optical depth on incidence angle is derived corresponds to an unrealistic isotropic phase function (Figs. 9c and 9d).

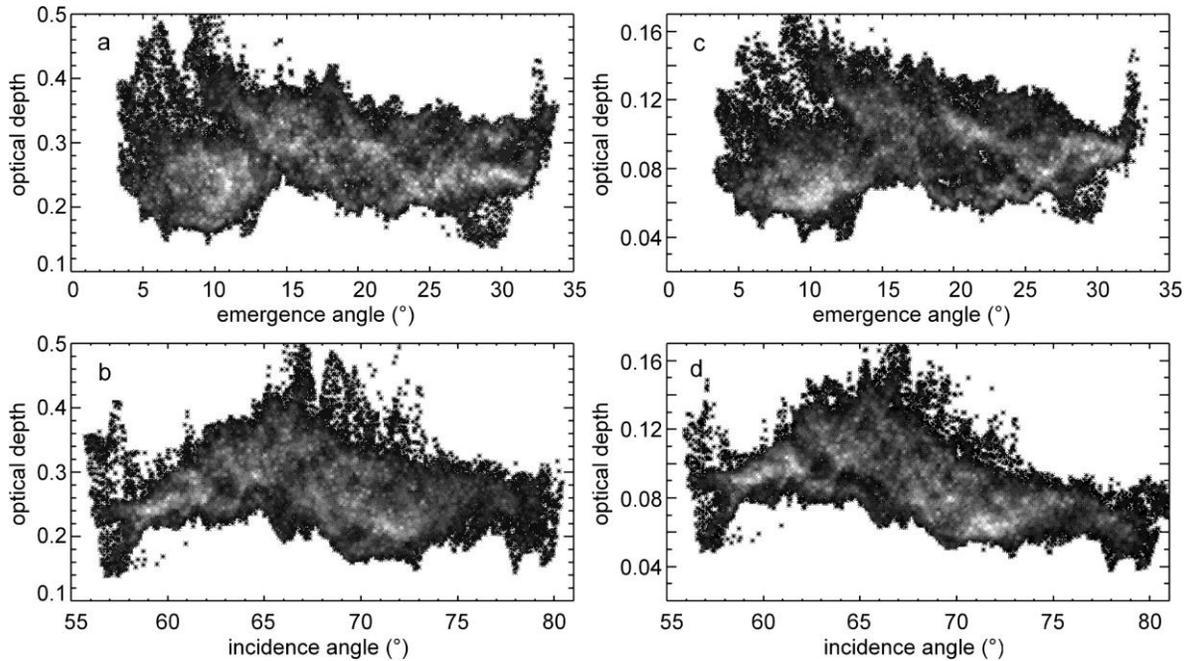

**Figure 9.** (top) *Correlation diagrams between the optical depth and a photometric angle for orbit No. 1781 ($L_s=224.5°$). The density of points ranges from black to white. (a and b) Aerosols parameters used in this study. (c and d) Isotropic phase function with w=0.932. (a and c) Emergence angle; (b and d) incidence angle. No correlations are observed between the photometric angles and the optical depth for our set of aerosols parameters. A possible correlation between the optical depth and the incidence angle is only noted if an isotropic phase function is used (see (d)).*

## 4. Discussion

We present in Fig. 10, Fig. 11 and Fig. 12 the mapping of the optical depth of aerosols above clean areas of the south polar cap as a function of $L_s$ with a time scale which is frequently ~0.5° of $L_s$ (~1 martian day). The extent of the useful regions for our optical depth retrieval method decreases with $L_s$ during the retreat of the seasonal cap. The optical depth above the perennial cap has been derived from $L_s=185°$ to $L_s=340°$ (Fig. 13a). Significant geographical and temporal variations of the optical depth are observed.

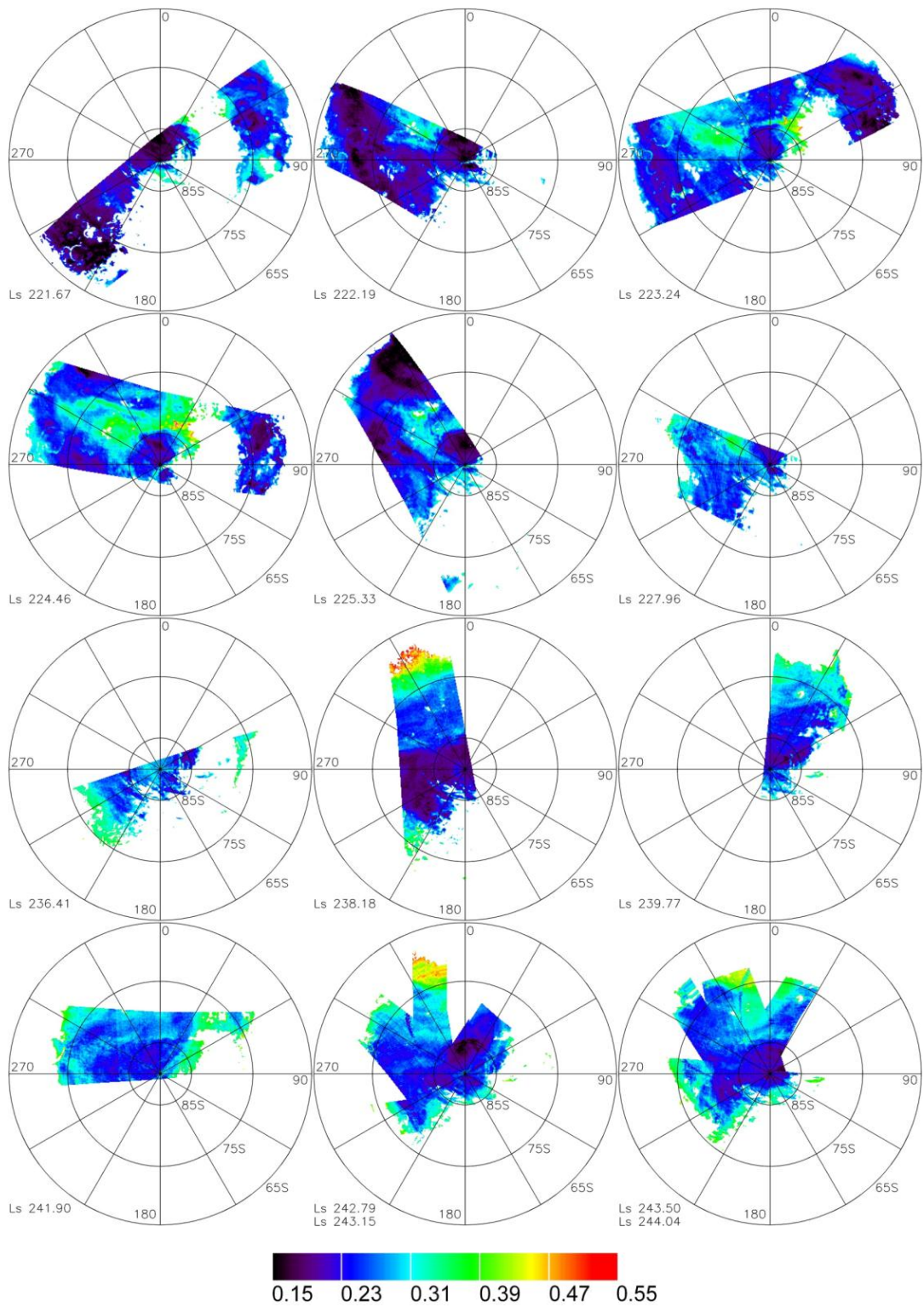

**Figure 10.** *Maps of the total optical depth of aerosols at 2.64 μm from $L_s$=221.7° to $L_s$=244.0°.*

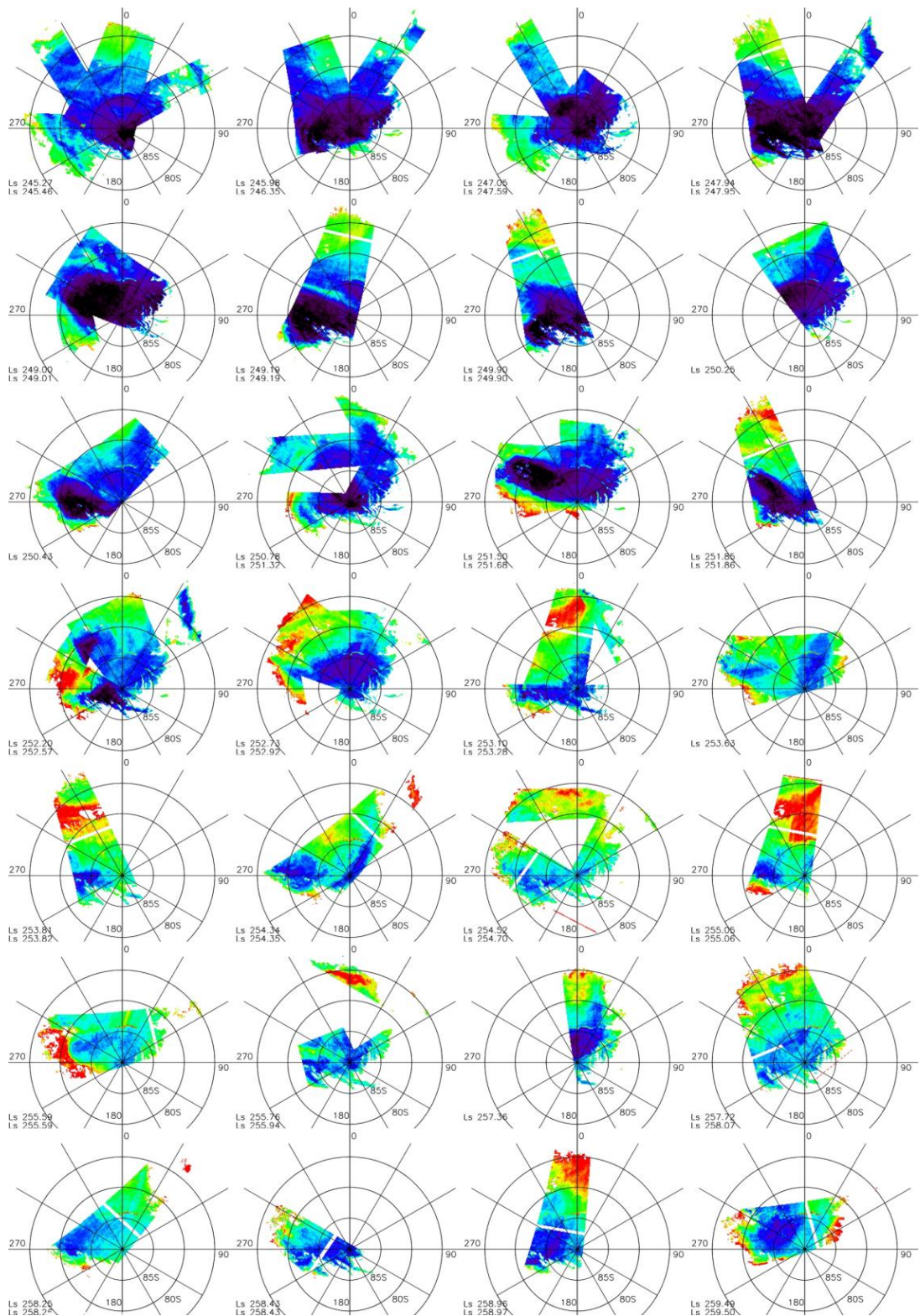

**Figure 11.** *Maps of the total optical depth of aerosols at 2.64 μm from Ls=245.3° to Ls=259.5°. The color code is the same as Fig. 10.*

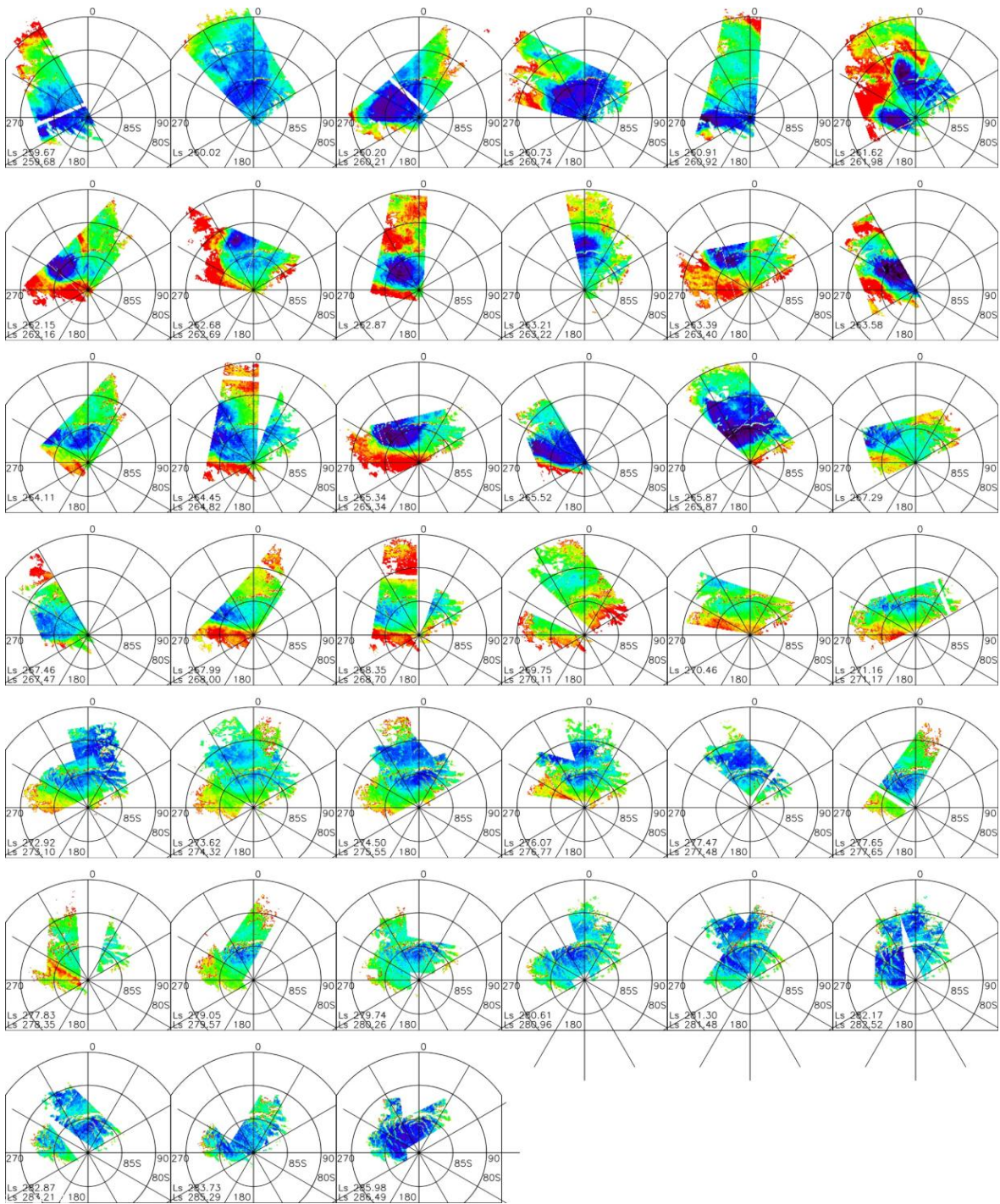

**Figure 12.** *Maps of the total optical depth of aerosols at 2.64 μm from Ls=259.7° to Ls=286.5°. The color code is the same as Fig. 10.*

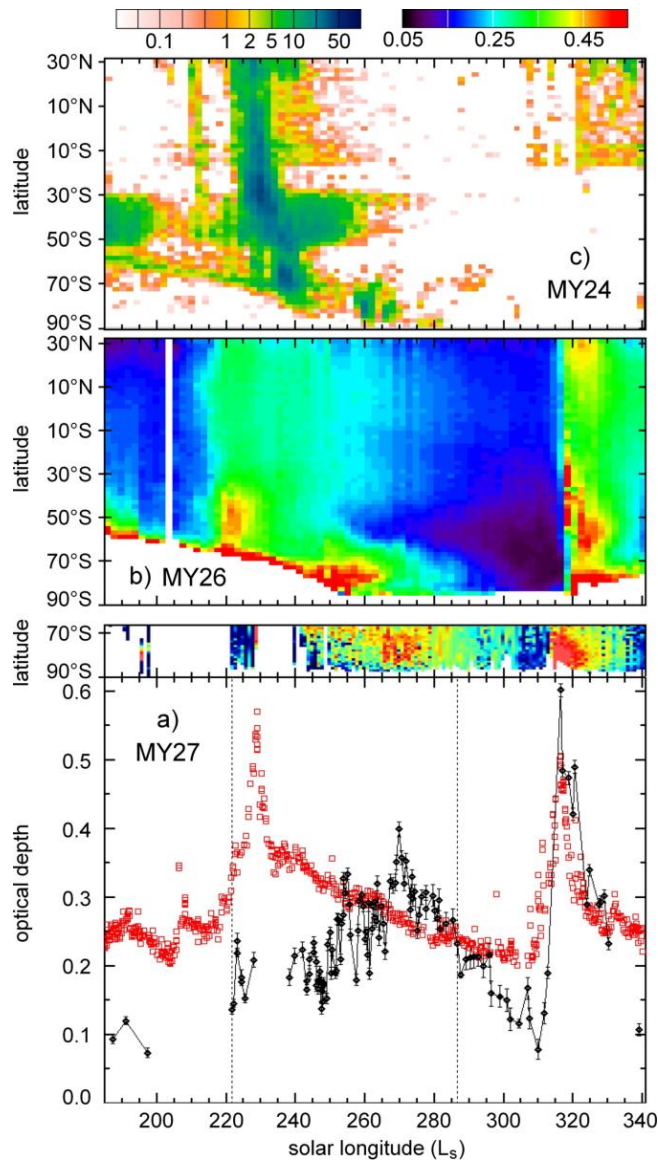

**Figure 13.** *Comparison with other datasets from the equinox (Ls=185°) to the end of summer (Ls=340°). (a) Black diamonds: optical depths determined from OMEGA observations above the perennial cap (87° S, 355° E) during MY 27; the error bars in terms of relative uncertainties due to the noise of the instrument have been evaluated as discussed in Section 3.3; red squares: optical depth from direct sunlight extinction measured by Pancam/Opportunity at 0.88 µm (Lemmon et al., 2006) during MY 27, scaled by a factor 0.3; (top) SPICAM optical depth in the UV at 6.1 mbar (Perrier et al., 2006) during MY 27 (the color code from blue to red corresponds to optical depths from 0 to 1). (b) TES dust optical depth at 9.3 µm and 6.1 mbar during MY 26 (Smith, 2006). (c) MOLA absorptive cloud density (%) during MY 24 (Neumann et al., 2003 G.A. Neumann, D.E. Smith and M.T. Zuber, Two Mars years of clouds detected by the Mars Orbiter Laser Altimeter, J. Geophys. Res. 108 (E4) (2003) 5023.Neumann et al., 2003). The dotted lines indicate the range of Ls of the maps presented in Fig. 10, Fig. 11 and Fig. 12.*

## 4.1. Spatial and temporal variability

The average trend of the temporal evolution is a low optical depth between $L_s=180°$ and $L_s=250°$ ($\tau(2.6~\mu m)=0.1–0.2$, Fig. 10 and Fig. 13), an increase of atmospheric dust activity observed between $L_s=250°$ and $L_s=270°$ ($\tau(2.6~\mu m)=0.3–0.6$, Fig. 11 and Fig. 12), then a decrease up to $L_s=310°$ (Fig. 12 and Fig. 13), a strong and rapid increase of the optical depth between $L_s=310°$ and $L_s=320°$ (from 0.1 to 0.6, Fig. 13a) and finally a decrease up to $L_s=340°$ (Fig. 13a).

This temporal evolution over the south permanent cap (355° E, 87° S, altitude: 4.7 km) is compared in Fig. 13a to the aerosol optical depth at 0.88 µm which was simultaneously measured by the MER "Opportunity," located at 1.95° S, 354.5° E (Lemmon et al., 2004, 2006). The values measured by Opportunity are scaled by a factor of 0.5 so as to account for the difference in wavelength ($\tau(0.88~\mu m)=2\tau(2.65~\mu m)$, see Section 2.2) and by a factor 0.6 so as to account for the difference of 6.2 km in altitude with an assumed scale height of 11.5 km (Lemmon et al., 2004; Zazova et al., 2005). We also compared our evaluations with the observations of TES (Fig. 13b) during the previous martian year (2003–2004, martian year 26 using the numbering of Clancy et al., 2000). The optical depth of aerosols is not measured by TES at high southern latitudes before $L_s=250°$. The absorptive clouds observed by MOLA mainly correspond to atmospheric dust. The observations over south polar regions during martian year 24 (Neumann et al., 2003) are available both before and after $L_s=250°$ (Fig. 13c). The optical depths in the UV retrieved by Perrier et al. (2006) from simultaneous SPICAM nadir observations are also compared with OMEGA results (Fig. 13a). For the 250°–340° range of solar longitudes, the OMEGA, SPICAM, MER and TES datasets are fully consistent: the optical depth between $L_s\sim250°$ and $L_s\sim270°$ increases in the south polar region whereas it decreases at low latitudes; After $L_s=270°$ the optical depth at high southern latitudes decreases simultaneously with that of mid-latitudes, but it reaches lower levels. At $L_s\sim315°$ the optical depth strongly increases over large areas of the planet including high-southern latitudes. We observed an increased optical depth of atmospheric dust at the edge of the seasonal cap during the retreat (Fig. 10, Fig. 11 and Fig. 12). This is consistent with both TES measurements and with the observations of absorptive clouds by MOLA near the edge of the seasonal cap (Fig. 13). Cantor et al. (2001) reported dust storms observed by MOC at the edge of the cap at this period. Kieffer et al. (2000) also observed dust clouds at the edge of the cap. The increase of optical depth at mid-latitudes at $L_s\sim220°–230°$ reported by the MERs, MOLA and TES is not observed by OMEGA at high southern latitudes. This is also consistent with MOLA observations.

Strong variations of the optical depth at relatively small spatial and temporal scales are observed by OMEGA. A dust cloud that changes in form and position is observed between $L_s=222°$ and $L_s=226°$ above the seasonal cap (Fig. 14). These results are consistent with those of Kieffer et al. (2000): transient dust clouds extending over a few hundred kilometers have been observed by TES in mid-spring. The temporal evolution of the optical depth for three representative regions monitored by OMEGA is summarized in Fig. 15. Rapid variations are observed which are specific to a given location. A zoom on a region that undergoes a major decrease of the optical depth in only one day is shown in Fig. 16a. The significant increase in albedo and band strength of observed spectra (Fig. 16b) requires a strong reduction of the optical depth of aerosols (from 0.7 to 0.2). The surface reflectance spectra which are determined by removing the aerosol contribution using our model are remarkably similar. Large spatial variations of the optical depth are observed over scales of a few tens of kilometers (Fig. 16a).

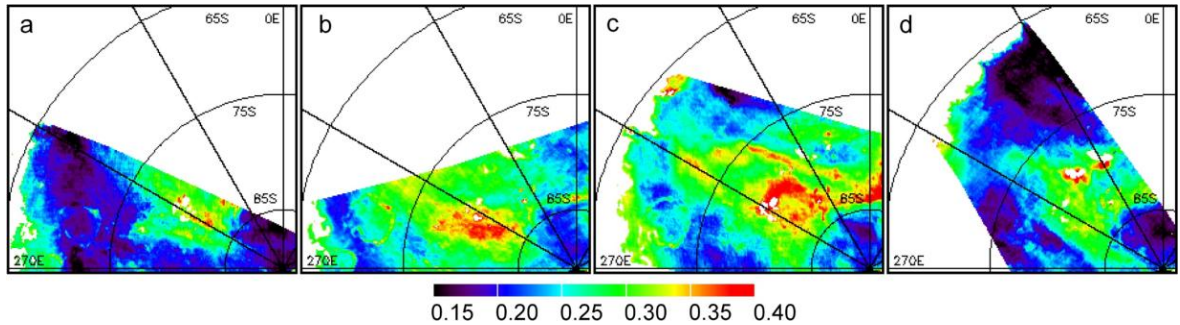

**Figure 14.** *Observations of a transient dust cloud (zoom from Fig. 10). Four maps of the optical depth of dust aerosols at 2.6 µm of the same area have been obtained with ~2 martian days between each (a, $L_s$=222.2°; b, $L_s$=223.2°; c, $L_s$=224.5°; d, $L_s$=225.3°). Variations of the optical depth by more than a factor of 2 are observed in this short time interval.*

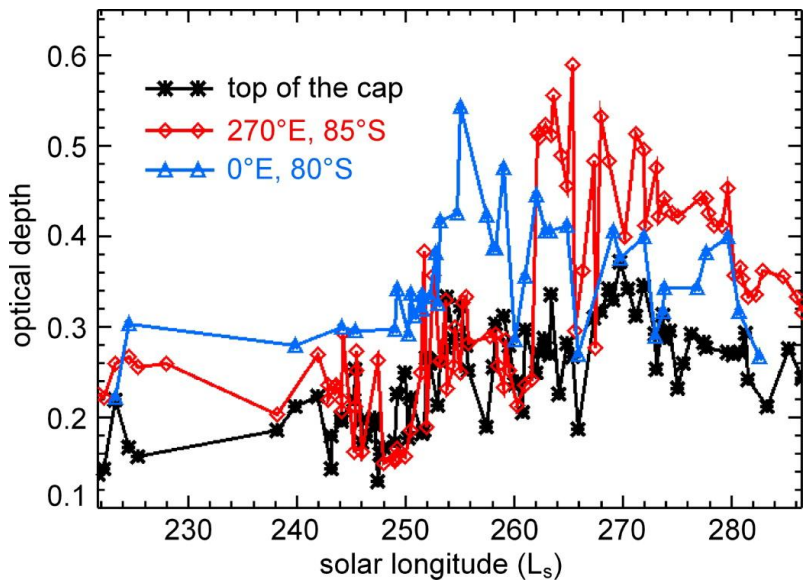

**Figure 15.** *Variations of the optical depth of aerosols for three regions: the highest region of the perennial cap, at 0° E, 87° S (black stars) and two long lasting regions of the seasonal cap, at 270° E, 85° S (red diamonds) and a region at 0° E, 80° S (blue triangles).*

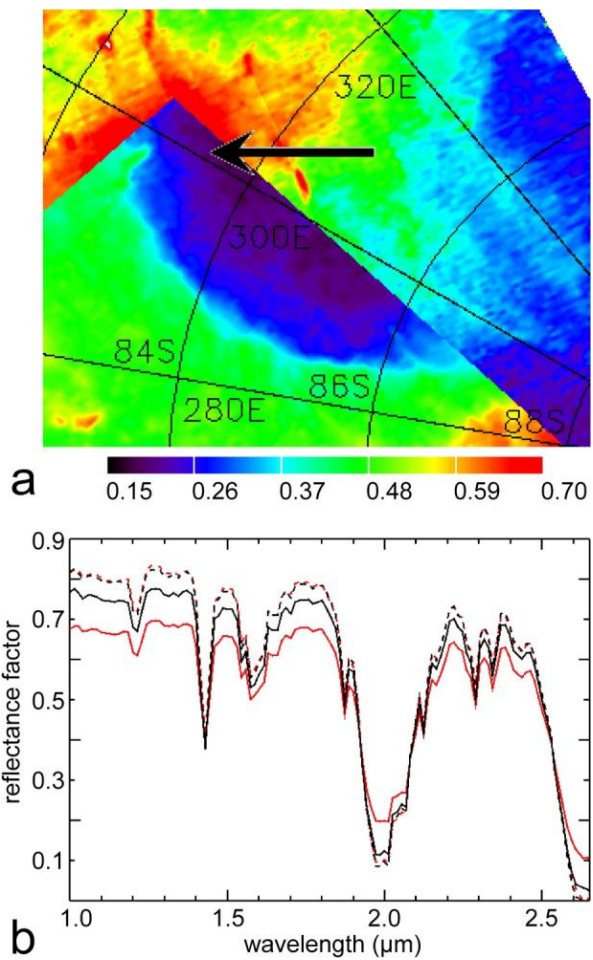

**Figure 16.** *(a) Map of the optical depth at $L_s=262.2°$ overlapping a map at $L_s=261.6°$ (approximately one day between the two observations). Both observations have been obtained with an incidence angle of ~60°. The arrow indicates a region (300.9° E, 83.5° S) where the optical depth decreases from 0.73 (red) to 0.21 (purple) in one day. (b) Near-IR observed spectra (solid lines) of the selected region at $L_s=261.6°$ (red) and at $L_s=262.2°$ (black). The spectra of the surface after removal of aerosol effects (dashed lines) are nearly identical.*

### 4.2. Correlation with altimetry

The variations of the optical depth with the elevation of the surface provide clues on the distribution of dust as a function of altitude. A correlation diagram for an observation during the period of strong dust activity ($L_s=250°–270°$) is presented in Fig. 17. It demonstrates that there are two components of aerosol loading at that time. Low optical depth variations can be modeled with a well-mixed dust component with a scale height of 11.5 km determined at low latitudes by Lemmon et al. (2004) at $L_s=14°$ and Zazova et al. (2005) at $L_s=338°$. TES measurements indicate that the atmospheric temperature is similar at high and low southern latitudes during spring and summer (Smith, 2006), so that the same scale height can be used. A component providing higher dust loading corresponds to localized increases of the dust opacity (dust cloud/storm), notably at the edge of the cap.

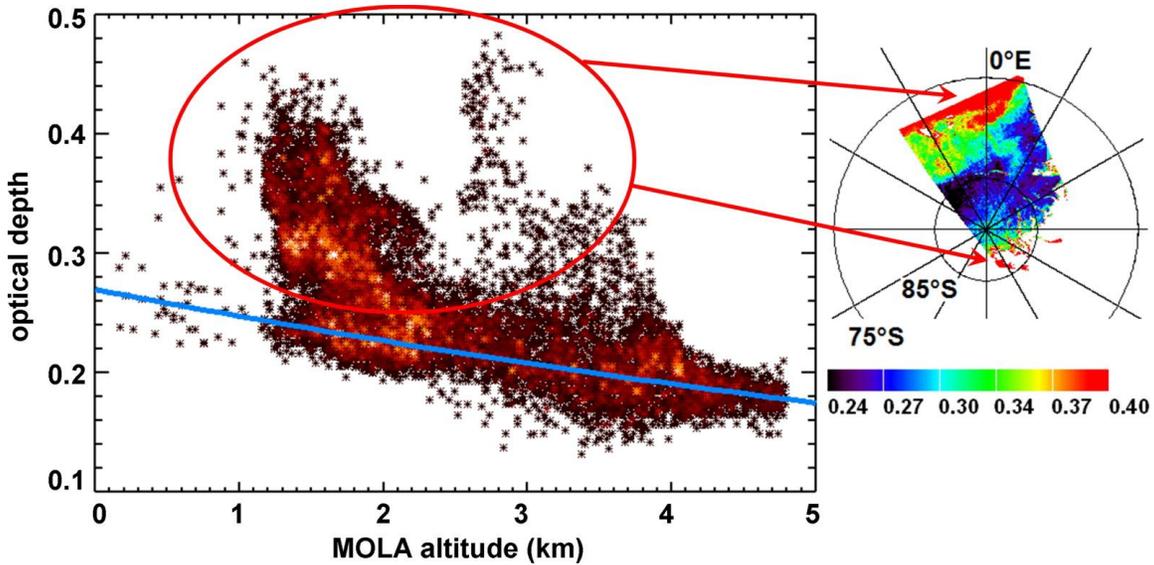

**Figure 17.** *(Left) Optical depth of aerosols at 2.64 μm as a function of altitude at $L_s=250.3°$. The density of points ranges from black through red to light orange. The blue solid line is the function of elevation $\tau=\tau(0)\exp(-h/11.5)$ which corresponds to well-mixed conditions. (Right) Map of the optical depth of aerosols scaled to an altitude of 0 km with the function indicated by the green line. Dust opacities exceeding that resulting from a well-mixed component are observed at the edge of the cap.*

At $L_s=220°–230°$, the optical depth over outer regions of the seasonal cap (latitude ~70° S) is lower than that expected from well-mixed conditions (Fig. 18a). Water ice clouds signatures in the 3 μm band are observed in these regions at $L_s=183°–193°$ (Langevin et al., 2007 and Fig. 18b) and GCM models (Forget et al., 1999; Montmessin et al., 2004) predict high precipitation rates for water ice in this period. The most likely explanation for the unusually low optical depth of aerosols in these regions is dust scavenging by water ice, a process that has been proposed by previous authors for the atmosphere of Mars (Newman et al., 2002b): water ice condensation on dust grains can speed up sedimentation, reducing the optical depth of aerosols. Neumann et al. (2003) mention this process as a possible explanation for the low density of absorptive clouds near the pole. It could explain why the increase of atmospheric dust occurring at mid-latitudes does not propagate to the high southern latitudes in that period (Fig. 13). The observation by OMEGA of surface spectra that correspond to an intra-mixture of dust grains within $CO_2$ ice contaminated by $H_2O$ ice at $L_s=222°$ (see Fig. 6, blue spectrum and Langevin et al., 2007) provides supporting evidence for this dust scavenging process.

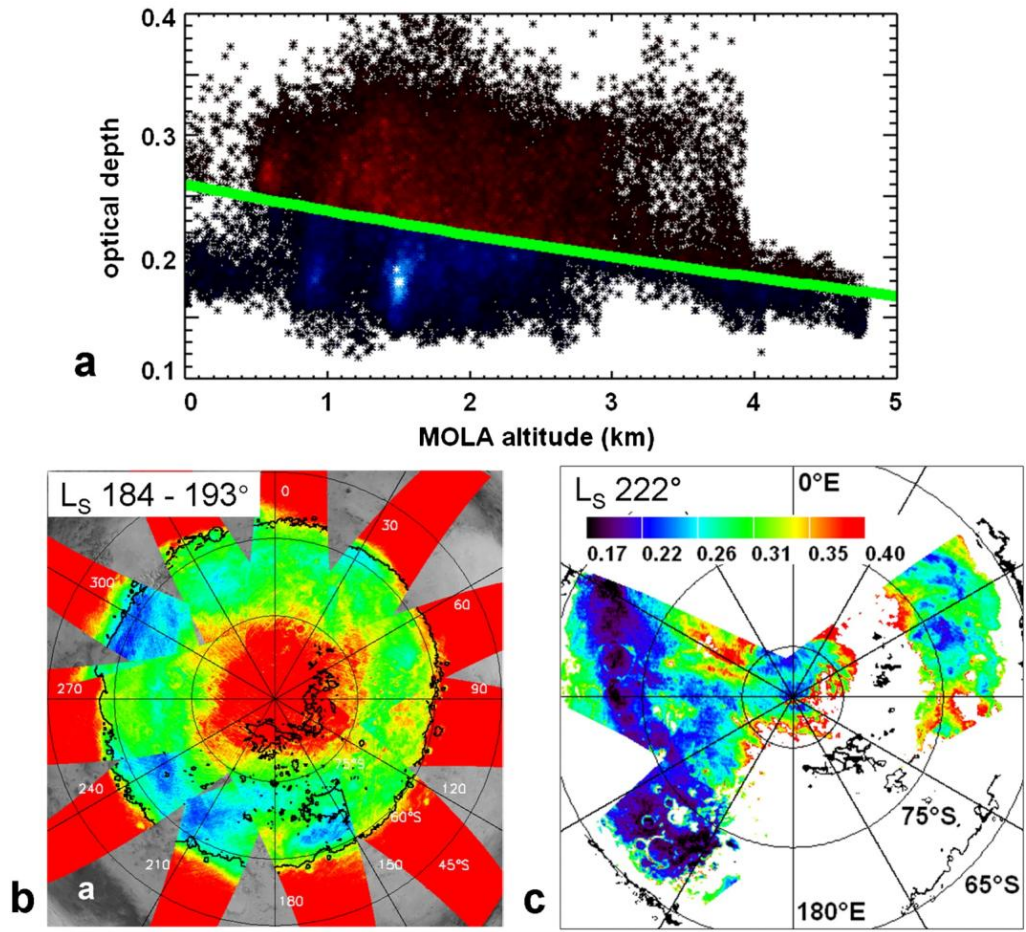

**Figure 18.** *(a) Correlation diagram between the optical depth of dust aerosols and the MOLA elevation at $L_s=222°$. The density of point ranges from dark to bright. The function $\tau=\tau(0)exp(-h/11.5)$ (green line) separates points in red with local high values of the optical depth (dust clouds) from points in blue with low values of the optical depth. (b) Map of water ice clouds at $L_s\sim190°$ from Langevin et al. (2007). (c) Map of the optical depth of dust aerosols at $L_s=222°$ scaled to an altitude of 0 km with the dependence indicated by the green line. Optical depths of dust aerosols lower than expected from well-mixed conditions (blue in (a) and (c)) are observed in the outer regions of the cap where water ice aerosols were observed at $L_s=190°$. The boundary of the cap is indicated with a black line in (b) and (c).*

### 4.3. Removal of atmospheric dust contribution in albedo maps

OMEGA albedo maps at 1.08 μm of the $CO_2$ ice cap (Langevin et al., 2007) are obtained by overlapping OMEGA tracks of the reflectance factor at 1.08 μm (continuum of the ice spectrum). Assuming no changes in the optical properties of the surface and a low impact of aerosols, overlapping tracks should be consistent if the surface is Lambertian. Consistent mosaics are obtained in most cases without correcting for surface photometric effects or aerosols effects. However, significant differences are occasionally observed (Langevin et al., 2007, Fig. 13). We have determined the optical depth of aerosols at 2.64 μm above ice-covered regions, and we can infer the optical depth at 1 μm from the relation $\tau(1\ \mu m)=1.9\times\tau(2.64\ \mu m)$ (see Section 2.2). We can therefore remove the contribution of aerosols at 1.08 μm by comparing the measured reflectance factors in the continuum with the look-up table of reflectance factors as a function of the Lambert albedo of the surface. The optical depth is not retrieve with our method above regions without clean ice. A first order removal of aerosols effects can however be performed above these regions by using the mean value of the optical depth that is inferred above the closest ice-covered regions. Maps obtained before and after the removal of aerosols effects are presented in Fig. 19 and Fig. 20.

This method provides satisfactory results above ice-covered regions (Fig. 19 and Fig. 20): the major discrepancies between overlapping tracks are indeed due to aerosols and they can be removed with our method. This result also confirms that the relation $\tau(1.08\ \mu m)=1.9\times\tau(2.64\ \mu m)$ characterizes dust aerosols above the cap during all the analyzed period. Results above ice-free regions are less convincing (some discrepancies between overlapping tracks increase). This is not surprising as we use for ice-free regions an optical depth propagated from the closest ice-covered region whereas Toigo et al. (2002) observe significant variations of the optical depth in the outer regions of the cap.

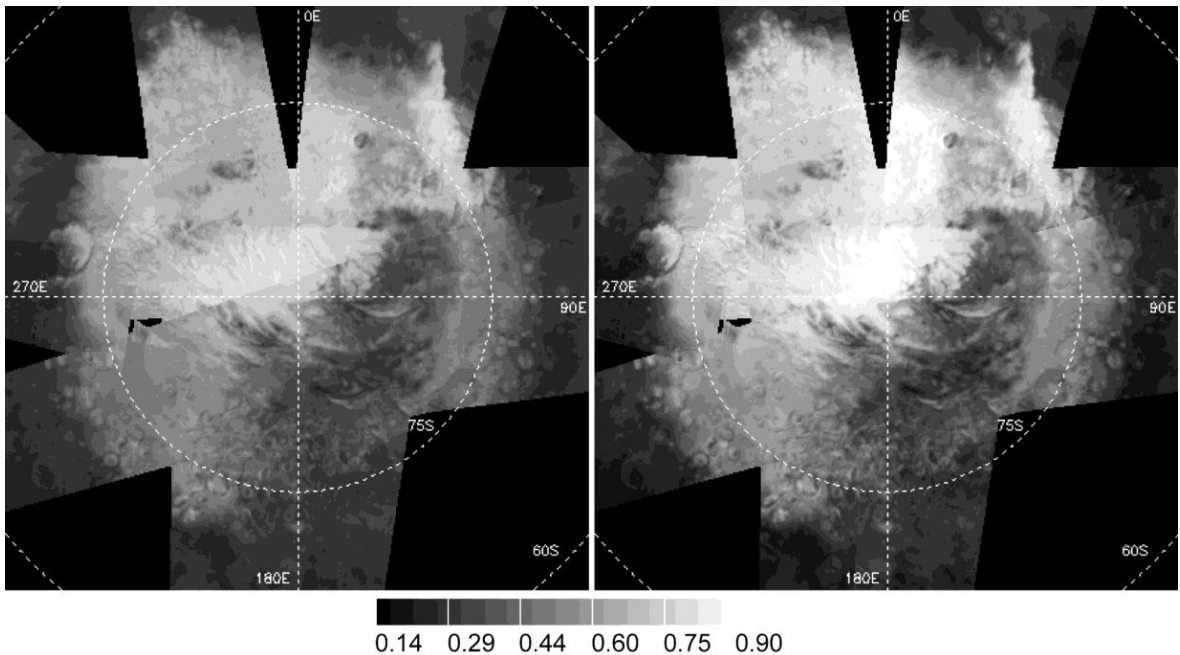

**Figure 19.** *Albedo map of the south seasonal cap of Mars at $L_s=236.4°–241.9°$. Left: albedo map without removal of aerosols effect. Right: same overlapping, but after removal of aerosols effect. The majority of the discrepancies between albedo levels is indeed due to aerosols, and can be remove using our model.*

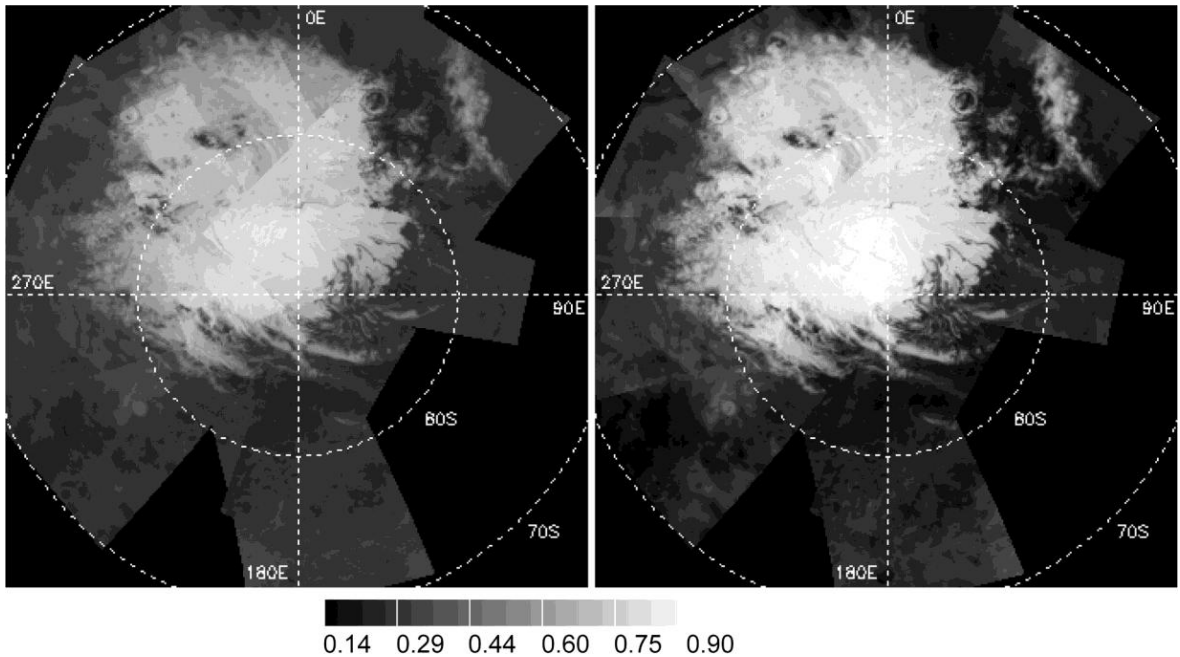

**Figure 20.** *Albedo map of the south seasonal cap of Mars at Ls=261.6°–262.7°. Left: albedo mosaic without removal of aerosols effect. Right: same mosaic, after removal of aerosols effect.*

5. **Conclusion**

A constant set of optical properties of dust aerosols is consistent with OMEGA observations of the south polar cap during southern spring and summer. Observations at different solar incidence angles and one EPF sequence have demonstrated that the reflectance in the 2.64 μm saturated absorption band of the surface $CO_2$ ice is mainly due to the light scattered by aerosols above a major fraction of the seasonal cap. This makes it possible to determine the optical depth of atmospheric dust over ice-covered regions when the ice is not contaminated by surface dust.

We have derived from the OMEGA dataset obtained in 2005 the optical depth of dust aerosols at 2.64 μm above a large fraction of the southern seasonal cap of Mars. Maps with a temporal scale ranging from one day to one week have been obtained between $L_s$=220° and $L_s$=290°. The optical depth of atmospheric dust has been determined above the permanent cap over a longer time range ($L_s$=185°–340°) covering the period from early spring to late summer of martian year 27.

The mean optical depth of dust aerosols at high southern latitudes increases by a factor of 3 during spring then decreases during early summer until the onset of a large dust storm at $L_s$ 315° which triggers a sharp increase in optical depth. Strong variations of the optical depth of dust aerosols are observed within a few ten kilometers and a few martian days over the seasonal cap. These localized dust clouds are most often observed near the edge of the seasonal cap. Evidence for dust scavenging by water ice in early spring has also been obtained. The OMEGA results in terms of spatial and temporal variability of aerosols are consistent with that provided by other instruments on recent missions: PanCam/Opportunity (Lemmon et al., 2004, 2006), TES/MGS (Kieffer et al., 2000; Smith, 2004, 2006), MOLA/MGS (Neumann et al., 2003), MOC/MGS (Cantor et al., 2001) and SPICAM/MEx (Perrier et al., 2006).

The determination of aerosol optical depths makes it possible to generate surface albedo maps corrected from the aerosol contributions which exhibit a high level of consistency between overlapping tracks with different lighting conditions. Such aerosol correction methods are useful for the interpretation of the near-IR spectral imaging observations of the surface of Mars by OMEGA and CRISM.

**References:**

Bibring et al., 2004 J.-P. Bibring and 13 colleagues, Perennial water ice identified in the south polar cap of Mars, *Nature* **428** (6983) (2004), pp. 627–630.

Cantor et al., 2001 B.A. Cantor, P.B. James, M. Caplinger and M.J. Wolff, Martian dust storms: 1999 Mars Orbiter Camera observations, *J. Geophys. Res.* **106** (E10) (2001), pp. 23653–23688.

Chassefiere et al., 1995 E. Chassefiere, P. Drossart and O. Korablev, Post-Phobos model for the altitude and size distribution of dust in the low martian atmosphere, *J. Geophys. Res.* **100** (E3) (1995), pp. 5525–5539.

Clancy et al., 2000 R.T. Clancy, B.J. Sandor, M.J. Wolff, P.R Christensen, M.D. Smith, J.C. Pearl, B.J. Conrath and R.J. Wilson, An intercomparison of ground-based millimeter, MGS TES, and Viking atmospheric temperature measurements: Seasonal and interannual variability of temperatures and dust loading in the global mars atmosphere, *J. Geophys. Res.* **105** (E4) (2000), pp. 9553–9571.

Clancy et al., 2003 R.T. Clancy, M.J. Wolff and P.R. Christensen, Mars aerosol studies with the MGS TES emission phase function observations: Optical depths, particle sizes, and ice cloud types versus latitude and solar longitude, *J. Geophys. Res.* **108** (E9) (2003) 5098.

Douté et al., 2007 S. Douté, B. Schmitt, Y. Langevin, F. Bibring, J.-P. Altieri, G. Bellucci, B. Gondet and F. Poulet, South pole of Mars: Nature and composition of the icy terrains from Mars Express OMEGA observations, *Planet. Space Sci.* **55** (2007), pp. 113–133.

Drossart et al., 1991 P. Drossart, J. Rosenqvist, S. Erard, Y. Langevin, J.-P. Bibring and M. Combes, Martian aerosol properties from the Phobos/ISM experiment, *Ann. Geophys.* **9** (1991), pp. 754–760.

Encrenaz et al., 2005 T. Encrenaz and 10 colleagues, A mapping of martian water sublimation during early northern summer using OMEGA/Mars Express, *Astron. Astrophys.* **441** (3) (2005), pp. L9–L12.

Erard et al., 1994 S. Erard, J. Mustard, S. Murchie, J.-P. Bibring, P. Cerroni and A. Coradini, Martian aerosols: Near-IR spectral properties and effects on the observation of the surface, *Icarus* **111** (1994), pp. 317–337. Abstract

Evans, 1998 K.F. Evans, The spherical harmonic discrete ordinate method for three-dimensional atmospheric radiative transfer, *J. Atmos. Sci.* **55** (1998), pp. 429–446. **Full Text** via CrossRef | View Record in Scopus | Cited By in Scopus (195)

Forget et al., 1999 F. Forget, F. Hourdin, R. Fournier, C. Hourdin, O. Talagrand, M. Collins, S.R. Lewis, P.L. Read and J.-P. Huot, Improved general circulation models of the martian atmosphere from the surface to above 80 km, *J. Geophys. Res.* **104** (E10) (1999), pp. 24155–24175.

Forget et al., 2008 Forget, F., Dolla, B., Vinatier, S., Spiga, A., 2008. A very simple algorithm to compute light scattering in optically thin planetary atmosphere. Application to remote sensing on Mars. Geophys. Res. Lett., submitted for publication.

Gooding, 1986 J.L. Gooding, Martian dust particles as condensation nuclei—A preliminary assessment of mineralogical factors, *Icarus* **66** (1986), pp. 56–74.

Kahre et al., 2006 M.A. Kahre, J.R. Murphy and R.M. Haberle, Modeling the martian dust cycle and surface dust reservoirs with the NASA Ames general circulation model, *J. Geophys. Res.* **111** (E6) (2006) E06008.

Kieffer, 1990 H.H. Kieffer, $H_2O$ grain size and the amount of dust in Mars' residual north polar cap, *J. Geophys. Res.* **95** (B2) (1990), pp. 1481–1493.

Kieffer et al., 2000 H.H. Kieffer, T.N. Titus, K.F. Mullins and P.R. Christensen, Mars south polar spring and summer behavior observed by TES: Seasonal cap evolution controlled by frost grain size, *J. Geophys. Res.* **105** (2000), pp. 9653–9700.

Kieffer et al., 2006 H.H. Kieffer, P.R. Christensen and T.N. Titus, $CO_2$ jets formed by sublimation beneath translucent slab ice in Mars' seasonal south polar ice cap, *Nature* **442** (2006), pp. 793–796.

Korablev et al., 1993 O.I. Korablev, V.A. Krasnopolsky, A.V. Rodin and E. Chassefiere, Vertical structure of martian dust measured by solar infrared occultations from the PHOBOS spacecraft, *Icarus* **102** (1) (1993), pp. 76–87.

Korablev et al., 2005 O. Korablev, V.I. Moroz, E.V. Petrova and A.V. Rodin, Optical properties of dust and the opacity of the martian atmosphere, *Adv. Space Res.* **35** (1) (2005), pp. 21–30.

Langevin et al., 2005 Y. Langevin, F. Poulet, J.-P. Bibring, B. Schmitt, S. Douté and B. Gondet, Summer evolution of the north polar cap of Mars as observed by OMEGA/Mars Express, *Science* **307** (2005), pp. 1581–1584

Langevin et al., 2006 Y. Langevin, S. Douté, M. Vincendon, F. Poulet, J.-P. Bibring, B. Gondet, B. Schmitt and F. Forget, No signature of clear $CO_2$ ice from the 'cryptic' regions in Mars' south seasonal polar cap, *Nature* **442** (2006), pp. 831–835.

Langevin et al., 2007 Y. Langevin, J.-P. Bibring, F. Montmessin, F. Forget, M. Vincendon, S. Douté, F. Poulet and B. Gondet, Observations of the south seasonal cap of Mars during retreat in 2004–2006 by the OMEGA visible/NIR imaging spectrometer on board Mars Express, *J. Geophys. Res.* **112** (2007) E08S12.

Lemmon et al., 2004 M.T. Lemmon and 14 colleagues, Atmospheric imaging results from the Mars Exploration Rovers: Spirit and Opportunity, *Science* **306** (2004), pp. 1753–1756.

Lemmon et al., 2006 M.T Lemmon and the Athena Science Team, Mars Exploration Rover atmospheric imaging: Dust storms, dust devils, dust everywhere, *Lunar Planet. Sci.* **XXXVII** (2006) Abstract 2181.

Mishchenko et al., 1996 M.I. Mishchenko, L.D. Travis and D.W. Mackowski, T-matrix computations of light scattering by nonspherical particles: A review, *J. Quant. Spectrosc. Radiat. Trans.* **55** (1996), pp. 535–575.

Montmessin et al., 2002 F. Montmessin, P. Rannou and M. Cabane, New insights into martian dust distribution and water-ice cloud microphysics, *J. Geophys. Res.* **107** (E6) (2002) 4-1.

Montmessin et al., 2004 F. Montmessin, F. Forget, P. Rannou, M. Cabane and R.M. Haberle, Origin and role of water ice clouds in the martian water cycle as inferred from a general circulation model, *J. Geophys. Res.* **109** (2004) E10004.

Montmessin et al., 2006 F. Montmessin, E. Quémerais, J.L. Bertaux, O. Korablev, P. Rannou and S. Lebonnois, Stellar occultations at UV wavelengths by the SPICAM instrument: Retrieval and analysis of martian haze profiles, *J. Geophys. Res.* **111** (2006) E09S09.

Neumann et al., 2003 G.A. Neumann, D.E. Smith and M.T. Zuber, Two Mars years of clouds detected by the Mars Orbiter Laser Altimeter, *J. Geophys. Res.* **108** (E4) (2003) 5023.

Newman et al., 2002a C.E. Newman, S.R. Lewis, P.L. Read and F. Forget, Modeling the martian dust cycle. 1. Representations of dust transport processes, *J. Geophys. Res.* **107** (E12) (2002) 6-1.

Newman et al., 2002b C.E. Newman, S.R. Lewis, P.L. Read and F. Forget, Modeling the martian dust cycle. 2. Multiannual radiatively active dust transport simulations, *J. Geophys. Res.* **107** (E12) (2002) 7-1.

Ockert-Bell et al., 1997 M.E. Ockert-Bell, J.F. Bell III, J.B. Pollack, C.P. McKay and F. Forget, Absorption and scattering properties of the martian dust in the solar wavelengths, *J. Geophys. Res.* **102** (E4) (1997), pp. 9039–9050

Perrier et al., 2006 S. Perrier, J.-L. Bertaux, F. Lefèvre, S. Lebonnois, O. Korablev, A. Fedorova and F. Montmessin, Global distribution of total ozone on Mars from SPICAM/MEX UV measurements, *J. Geophys. Res.* **111** (2006) E09S06.

Poulet et al., 2002 F. Poulet, J.N. Cuzzi, D.P. Cruikshank, T. Roush and C.M. Dalle Ore, Comparison between the Shkuratov and Hapke scattering theories for solid planetary surfaces. Application to the surface composition of two centaurs, *Icarus* **160** (2) (2002), pp. 313–324.

Shkuratov et al., 1999 Y. Shkuratov, Y. Starukhina, H. Hoffmann and G. Arnold, A model of spectral albedo of particulate surfaces: Implications for optical properties of the Moon, *Icarus* **137** (1999), pp. 235–246.

Smith, 2004 M.D. Smith, Interannual variability in TES atmospheric observations of Mars during 1999–2003, *Icarus* **167** (1) (2004), pp. 148–165.

Smith, 2006 Smith, M.D., 2006. TES atmospheric temperature, aerosol optical depth, and water vapor observations 1999–2004. Paper presented at the second workshop on Mars atmosphere modeling and observations, Granada, Spain.

Toigo et al., 2002 A.D. Toigo, M.I. Richardson, R.J. Wilson, H. Wang and A.P. Ingersoll, A first look at dust lifting and dust storms near the south pole of Mars with a mesoscale model, *J. Geophys. Res.* **107** (E7) (2002) 4-1.

Tomasko et al., 1999 M.G. Tomasko, L.R. Doose, M. Lemmon, P.H. Smith and E. Wegryn, Properties of dust in the martian atmosphere from the Imager on Mars Pathfinder, *J. Geophys. Res.* **104** (E4) (1999), pp. 8987–9008.

Vincendon et al., 2007a M. Vincendon, Y. Langevin, F. Poulet, J.-P. Bibring and B. Gondet, Recovery of surface reflectance spectra and evaluation of the optical depth of aerosols in the near-IR using a Monte-Carlo approach: Application to the OMEGA observations of high latitude regions of Mars, *J. Geophys. Res.* **112** (2007) E08S13.

Vincendon et al., 2007b M. Vincendon, Y. Langevin, F. Poulet, J.-P. Bibring and B. Gondet, Retrieval of surface Lambert albedos and aerosols optical depths using OMEGA near-IR EPF observations of Mars, *Lunar Planet. Sci.* **XXXVIII** (2007) Abstract 1650.

Wolff and Clancy, 2003 M.J. Wolff and R.T. Clancy, Constraints on the size of martian aerosols from Thermal Emission Spectrometer observations, *J. Geophys. Res.* **108** (E9) (2003) 5097.

Wolff et al., 2006 M.J. Wolff and 12 colleagues, Constraints on dust aerosols from the Mars Exploration Rovers using MGS overflights and Mini-TES, *J. Geophys. Res.* **111** (2006) E12S17.

Zazova et al., 2005 L. Zazova and 20 colleagues, Water clouds and dust aerosols observations with PFS MEX at Mars, *Planet. Space Sci.* **53** (10) (2005), pp. 1065–1077.